\documentclass[journal]{IEEEtran}

\usepackage{xcolor,soul,framed}
\colorlet{shadecolor}{yellow}
\usepackage[pdftex]{graphicx}
\graphicspath{{outputs/causal_isolation/causal_isolation_suite_latest/}}
\DeclareGraphicsExtensions{.pdf,.png,.jpg,.jpeg}

\usepackage[cmex10]{amsmath}
\usepackage{amssymb}
\usepackage{amsthm}
\usepackage{bm}
\usepackage{array}
\usepackage{url}
\usepackage{cite}
\usepackage[caption=false,font=footnotesize]{subfig}
\usepackage{tikz}
\usetikzlibrary{arrows.meta,positioning}

\newcommand{\tr}{\operatorname{tr}}
\newcommand{\E}{\mathbb{E}}
\newcommand{\kappaV}{\kappa(V)}
\newcommand{\Acl}{A_{\mathrm{cl}}}
\newcommand{\Weff}{W_{\mathrm{eff}}}

\begin{document}

\title{Input-Side Variance Suppression under Non-Normal Transient Amplification in Continuous-Control Reinforcement Learning}

\author{Yue~Wu%
\thanks{Y. Wu is with the School of Automation, Xi'an Jiaotong University, Xi'an 710049, China, and also with Xinjiang Cigarette Factory, Hongyun Honghe Tobacco (Group) Co., Ltd., Urumqi 830000, China (e-mail: wuyue0619@stu.xjtu.edu.cn).}}

\markboth{IEEE CONTROL SYSTEMS LETTERS,~VOL.~X, NO.~X,~2026}%
{Wu: Input-Side Variance Suppression under Non-Normal Transient Amplification}

\maketitle

\begin{abstract}
Continuous-control reinforcement learning (RL) often exhibits large closed-loop variance, high-frequency control jitter, and sensitivity to disturbance injection. Existing explanations usually emphasize disturbance sources such as action noise, exploration perturbations, or policy nonsmoothness. This letter studies a complementary amplifier-side perspective: in nominally stable yet strongly non-normal closed loops, small input perturbations can undergo transient amplification and lead to disproportionately large state covariance. Motivated by this source--amplifier decomposition, we introduce an input-side variance suppression layer that operates between the learned policy and the plant input to reduce applied-input variance and step-to-step jitter. To separate mechanism from correlation, we use two control-theoretic interventions: one varies only eigenvector geometry under fixed eigenvalues and spectral radius, and the other varies only applied-input statistics under fixed strongly non-normal geometry. We then provide mechanism-consistent external validation on planar quadrotor tasks. Throughout, Koopman/ALE surrogates are used only as analysis and certification tools, not as direct performance paths. Taken together, the results support a narrower claim: in the studied settings, non-normal transient amplification is an important and under-emphasized contributor to execution-time closed-loop variance, and source-side suppression can reduce downstream covariance without changing the structural peak gain.
\end{abstract}


\IEEEpeerreviewmaketitle

\section{Introduction}

\IEEEPARstart{C}{ontinuous}-control reinforcement learning (RL) has shown strong empirical performance in robotics and dynamical decision-making, yet learned controllers often exhibit large closed-loop jitter, high-frequency actuation, and elevated rollout variance at deployment time \cite{cheng_control_regularization,mysore_caps,lee_grad_caps,raffin_smooth_exploration,td3,sac}. Existing explanations typically attribute these effects to disturbance sources, including action noise, exploration perturbations, reward-design artifacts, function-approximation error, and policy nonsmoothness \cite{tessler_action_robust,parisi_td_regularized,dong_reward_smoothing,korenkevych_autoregressive,papini_smoothing}. This source-centered view has motivated action regularization, smooth exploration, and related methods that aim to reduce visibly jerky control signals \cite{mysore_caps,lee_grad_caps,raffin_smooth_exploration,cao_image_caps}.

However, a source-only explanation leaves an important question under-emphasized: why can small input-side perturbations still produce large downstream state covariance even when the closed loop is nominally stable? From a dynamical-systems perspective, nominal stability alone does not imply transiently benign behavior. In strongly non-normal closed loops, the geometry of the state-transition operator can transiently amplify disturbances despite a stable spectral radius \cite{trefethen1993,farrell1996,schmid2007,farrell1994}. Thus, similar source statistics can induce very different state variance depending on the closed-loop amplifier.

This letter adopts a source--amplifier decomposition of closed-loop variance in continuous-control RL. We treat action noise, exploration perturbations, and policy jitter as disturbance sources, while treating non-normal closed-loop geometry as an amplifier that shapes downstream disturbance propagation. Under this view, reducing closed-loop variance is not only a matter of smoothing the source; it is also a matter of understanding how disturbances are amplified after entering the loop. This perspective motivates an input-side variance suppression layer placed between the learned policy and the plant input. The layer does not modify the structural amplifier directly. Instead, it alters applied-input statistics so as to reduce disturbance injection into a strongly non-normal closed loop.

To support this mechanism, we use two control-theoretic interventions. In the first, we hold eigenvalues, spectral radius, input channel, and disturbance statistics fixed, and vary only eigenvector geometry to isolate non-normal amplification. In the second, we hold the strongly non-normal closed-loop geometry fixed, and vary only applied-input statistics to isolate source-side intervention. We then bridge these findings to a planar quadrotor study as mechanism-consistent external validation. Koopman/ALE surrogates remain analysis and certification branches rather than direct performance paths \cite{takase2020,cui2022,huang2019,sinha2022,narasingam2023,strasser2023,fan2024}. The central message is correspondingly narrow: in the studied settings, closed-loop variance is usefully analyzed through both disturbance sources and closed-loop amplifiers.

The main contributions are summarized as follows.
\begin{itemize}
    \item We introduce a source--amplifier interpretation of execution-time closed-loop variance and identify non-normality as an under-emphasized disturbance amplifier in nominally stable closed loops.
    \item We propose a simple post-training input-side suppressor that intervenes on applied-input statistics without changing the structural amplifier.
    \item We provide controlled-intervention evidence that isolates amplifier-side and source-side effects separately, complemented by mechanism-consistent planar-quadrotor validation while restricting Koopman/ALE surrogates to analysis and certification roles.
\end{itemize}

The paper is positioned as a mechanism-focused control letter centered on controlled mechanism isolation; head-to-head benchmarking against existing smooth-control RL baselines is left to future work.

\section{Problem View: Source versus Amplifier}

\subsection{What Variance Means in This Letter}

This letter studies execution-time closed-loop variance rather than training-time estimator variance. The quantities of interest are applied-input variance, state covariance, transient peak gain, and nominal stability. In contrast, policy-gradient variance reduction \cite{greensmith2004,qprop,xu_svrpg,wu_action_baseline,cheng_control_variates,liu_stein,zhao_policy_gradient,munos2006,wang2020} is not the primary object of study here.

We therefore distinguish structural amplification metrics from rollout-level statistics. Structural amplification is quantified by
\begin{equation}
G_{\mathrm{peak}} := \sup_{k \ge 0}\|\Acl^k\|_2,
\end{equation}
whereas rollout-level dispersion is quantified through the empirical covariance
\begin{equation}
\hat{\Sigma}_x := \frac{1}{T}\sum_{t=1}^{T}(x_t-\bar{x})(x_t-\bar{x})^\top
\end{equation}
and the expected peak state norm
\begin{equation}
J_{\mathrm{peak}} := \E\!\left[\max_{1 \le t \le T}\|x_t\|_2\right].
\end{equation}
This distinction matters for the rest of the paper: the suppressor changes applied-input statistics and therefore rollout-level disturbance injection, whereas the amplifier is identified through fixed structural properties of the closed loop.

\subsection{Minimal Closed-Loop Model}

We adopt the discrete-time linearized closed-loop model
\begin{equation}
x_{t+1} = \Acl x_t + G w_t,
\label{eq:closed_loop}
\end{equation}
where $\Acl$ denotes the local closed-loop transition matrix, $G$ is the disturbance channel, and $w_t$ is the applied input disturbance after source shaping and suppression. When $\rho(\Acl)<1$, the steady-state covariance exists and satisfies
\begin{equation}
\Sigma_x = \sum_{k=0}^{\infty} \Acl^k G \Weff G^\top (\Acl^k)^\top,
\label{eq:sigma_series}
\end{equation}
with $\Weff$ the effective disturbance covariance. Equation~\eqref{eq:sigma_series} already separates two roles: $\Weff$ is determined by the source-side pathway, while the geometry of $\Acl$ determines how strongly disturbances are propagated and transiently amplified.

In this letter, the model is used as a local analysis lens rather than as a claim of globally exact nonlinear dynamics. Its role is to make source-side disturbance shaping and amplifier-side propagation analytically distinguishable near the operating regimes studied in the experiments.

Nominal stability does not imply transient benignness. When $\Acl$ is strongly non-normal, the finite-time operator norm
\begin{equation}
G_{\mathrm{peak}} := \sup_{k \ge 0}\|\Acl^k\|_2
\label{eq:peak_gain}
\end{equation}
can be large even if $\rho(\Acl) < 1$ \cite{trefethen1993,schmid2007}. This is the amplifier effect that we make explicit.

\subsection{Source--Amplifier Causal Chain}

\begin{figure}[t]
\centering
\footnotesize
\begin{tikzpicture}[
    node distance=3.8mm,
    >=Latex,
    stage/.style={draw, rounded corners=1.4pt, align=center, inner sep=3pt, minimum width=0.76\linewidth, fill=white},
    analysis/.style={draw, rounded corners=1.4pt, align=center, inner sep=2.5pt, minimum width=0.56\linewidth, fill=white},
    main/.style={-Latex, line width=0.9pt},
    shaped/.style={-Latex, line width=0.9pt, dashed},
    inject/.style={-Latex, line width=0.9pt, densely dotted},
    amplify/.style={-Latex, line width=1.1pt},
    analyze/.style={-Latex, gray, line width=0.7pt, dashed}
]
\node[stage] (source) {\textbf{Source}\\ $\eta_t$\\ \scriptsize action noise / exploration / policy jitter};
\node[stage, below=of source] (suppressor) {\textbf{Suppressor}\\ $u_t^{\mathrm{app}} = F(u_t^\pi,\eta_t)$\\ \scriptsize input-side variance suppression};
\node[stage, below=of suppressor] (stats) {\textbf{Applied-input Statistics}\\ $\Weff$, PSD, temporal correlation};
\node[stage, below=of stats] (amp) {\textbf{Non-normal Amplifier}\\ $\Acl$, $\rho(\Acl)<1$, $\kappaV$};
\node[stage, below=of amp] (gain) {\textbf{Transient Gain}\\ $G_{\mathrm{peak}} = \sup_{k \ge 0}\|\Acl^k\|_2$};
\node[stage, below=of gain] (cov) {\textbf{Outcome: State Covariance}\\ $\tr(\Sigma_x)$, $J_{\mathrm{peak}}$};
\node[analysis, below=4mm of cov] (ale) {\textbf{Koopman/ALE}\\ \scriptsize diagnosis / certification only};

\draw[main] (source) -- node[right, align=left, font=\scriptsize] {disturbance\\ generation} (suppressor);
\draw[shaped] (suppressor) -- node[right, align=left, font=\scriptsize] {source shaping} (stats);
\draw[inject] (stats) -- node[right, align=left, font=\scriptsize] {injected\\ statistics} (amp);
\draw[amplify] (amp) -- node[right, align=left, font=\scriptsize] {non-normal\\ propagation} (gain);
\draw[amplify] (gain) -- node[right, align=left, font=\scriptsize] {variance\\ accumulation} (cov);
\path (amp.east) ++(11mm,0) coordinate (alebend);
\draw[analyze] (amp.east) -- (alebend) |- node[pos=0.3, right, font=\scriptsize, text=gray] {analysis only} (ale.north);
\end{tikzpicture}
\caption{Source--amplifier mechanism of closed-loop variance. Line styles distinguish source generation, source shaping, disturbance injection, and downstream amplification. Koopman/ALE is analysis-only.}
\label{fig:causal_chain}
\end{figure}
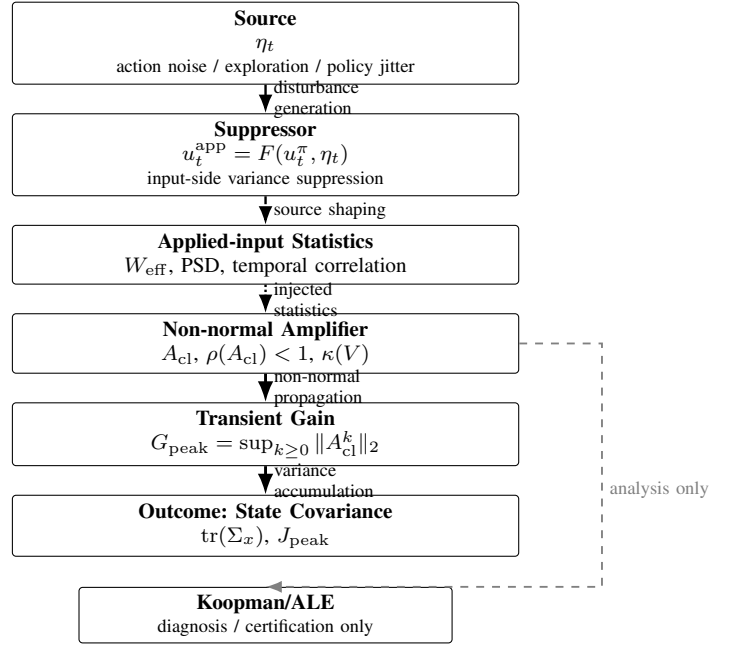

Fig.~\ref{fig:causal_chain} summarizes the mechanism studied in this letter. The source includes action noise, exploration perturbations, policy jitter, and other input-side disturbances. The suppressor is the proposed input-side variance suppression layer. The amplifier is the non-normal geometry of $\Acl$, and the downstream outcome is measured through $G_{\mathrm{peak}}$, $\tr(\Sigma_x)$, and the expected peak state norm.

\subsection{Intervention Statements}

\noindent\textbf{Amplifier-isolation statement.} If the eigenvalue set, spectral radius, input channel, and $\Weff$ are fixed, then changes in $G_{\mathrm{peak}}$ or $\tr(\Sigma_x)$ should be attributed to the non-normal geometry of $\Acl$ rather than to nominal instability.

\medskip
\noindent\textbf{Source-side intervention statement.} If $\Acl$ is fixed, then an input-side suppressor can affect $\tr(\Sigma_x)$ only by altering applied-input statistics or their spectrum; in particular, a source-side intervention need not change structural peak gain.

\medskip
These are not presented as new formal theorems; rather, they state the intervention logic used to organize the synthetic evidence and interpret the nonlinear bridge results.

\section{Input-Side Variance Suppression}

\subsection{Method Definition}

Let
\begin{equation}
u_t^\pi := \pi_\theta(x_t)
\end{equation}
denote the raw action generated by a trained policy at state $x_t$. We insert an input-side suppressor between the policy output and the plant input and denote the resulting applied control by $u_t^{\mathrm{app}}$. In the intervention view adopted in this letter, source-side perturbations are represented by $\eta_t$, and the suppressor maps the policy-side signal and the source perturbation to the applied signal seen by the plant:
\begin{equation}
u_t^{\mathrm{app}} = F_\beta\!\left(u_t^\pi,\eta_t,u_{t-1}^{\mathrm{app}}\right).
\end{equation}
Thus, the suppressor acts on the policy-to-plant input path. It changes the applied-input statistics seen by the plant, but it does not directly modify the structural closed-loop operator $A_{\mathrm{cl}}$.

To keep the mechanism explicit, we distinguish three objects throughout the section: the raw policy command $u_t^\pi$, the source-side perturbation $\eta_t$, and the applied plant input $u_t^{\mathrm{app}}$. In the synthetic interventions, white versus temporally filtered source realizations produce different applied-input statistics after passing through the suppressor. In the UAV bridge study, the same raw controller is retained while the source-side injection path is altered, so that the suppressor remains a source-side intervention rather than a controller redesign.

\subsection{Mechanism Alignment}

The role of the suppressor is best understood through the source--amplifier decomposition introduced in Section II. Under the local closed-loop model
\begin{equation}
x_{t+1}=A_{\mathrm{cl}}x_t + G w_t,
\end{equation}
the steady-state covariance is
\begin{equation}
\Sigma_x
=
\sum_{k=0}^{\infty}
A_{\mathrm{cl}}^k\, G\, W_{\mathrm{eff}}\, G^\top (A_{\mathrm{cl}}^k)^\top,
\end{equation}
where $W_{\mathrm{eff}}$ denotes the effective disturbance covariance after source shaping and suppression. The structural amplification of the closed loop is characterized by
\begin{equation}
G_{\mathrm{peak}}
:=
\sup_{k\ge 0}\|A_{\mathrm{cl}}^k\|_2.
\end{equation}

The proposed suppressor acts on the source/suppressor side of the chain by modifying $W_{\mathrm{eff}}$ or, more generally, the effective spectrum of the injected disturbance. It does not directly alter $A_{\mathrm{cl}}$ and therefore does not directly alter $G_{\mathrm{peak}}$. Consequently, under fixed $A_{\mathrm{cl}}$, any reduction in $\mathrm{tr}(\Sigma_x)$ should be interpreted as a source-side intervention effect rather than an amplifier redesign. This distinction is what allows CI-2 to separate covariance reduction from structural gain reduction.

\subsection{Suppressor Instantiation and Deployment}

For all reported experiments, we instantiate $F_\beta$ as a simple first-order causal temporal smoother. Define the policy-side signal
\begin{equation}
u_t^{\mathrm{raw}} := u_t^\pi + \eta_t,
\end{equation}
and update the applied signal according to
\begin{equation}
u_t^{\mathrm{app}}
=
(1-\beta)u_{t-1}^{\mathrm{app}} + \beta u_t^{\mathrm{raw}},
\qquad
\beta \in (0,1].
\end{equation}
Equivalently,
\begin{equation}
u_t^{\mathrm{app}} - u_{t-1}^{\mathrm{app}}
=
\beta\bigl(u_t^{\mathrm{raw}} - u_{t-1}^{\mathrm{app}}\bigr),
\end{equation}
which shows that $\beta$ governs the update rate of the applied command toward the current raw input. Smaller $\beta$ yields stronger smoothing and slower response, while $\beta=1$ recovers the unsuppressed input path.

In the reported experiments, we use a fixed $\beta=0.85$ across all intervention and UAV evaluations. This choice is deliberately lightweight: the goal of the letter is not to optimize a rich filter family or to tune a scenario-specific bandwidth, but to test whether a simple source-side intervention produces predictable downstream covariance changes under fixed amplifier geometry. Practically, $\beta=0.85$ serves as a conservative operating point that visibly suppresses applied-input variance in both the synthetic and UAV studies without qualitatively changing the nominal closed-loop behavior under the reported operating conditions. The applied-input trajectory is initialized by the first available command,
\begin{equation}
u_0^{\mathrm{app}} := u_0^{\mathrm{raw}},
\end{equation}
and the suppressor is then updated causally at every time step. The intervention is used within the same plant-input interface as the raw controller and is intended to preserve the nominal actuation pathway rather than redefine it.

A practical feature of the construction is that it is post-training compatible. The policy $\pi_\theta$ is trained first, and the suppressor is inserted only during deployment/evaluation. No retraining, policy-architecture modification, or controller-resynthesis step is required. This keeps the intervention aligned with the source side of the mechanism chain and avoids conflating suppressor effects with changes in the structural amplifier.

\subsection{Practical Scope and Non-Novelty of the Filter Structure}

The novelty of this letter is not the smoother structure itself, but the source--amplifier intervention view and the controlled-isolation evidence built around it. We deliberately use a first-order causal smoother because the purpose of the letter is mechanism isolation rather than filter-family optimization. The contribution is therefore not a new low-pass design, but an explicit control-theoretic account of how a simple source-side suppressor interacts with a non-normal closed-loop amplifier.

\subsection{Koopman/ALE as an Analysis Branch}

Koopman/ALE surrogates are used only to diagnose local propagation structure, nominal stability, and covariance-related certificates. They support analysis and certification-oriented comparisons, but they are not presented as the mechanism of empirical variance reduction. In particular, projection and contractive certification belong to the analysis branch rather than the direct physical performance path. Accordingly, any observed reduction in applied-input variance or state covariance is attributed to the source-side intervention above, not to spectral projection itself.

\section{Controlled Interventions and UAV Validation}

\subsection{Intervention Protocol and Measured Quantities}

Table~\ref{tab:protocol} summarizes the three evidence blocks used in this paper. CI-1 and CI-2 are synthetic interventions designed for mechanism isolation. CI-3 is a bridge experiment intended to assess whether the same source--amplifier interpretation remains informative in a nonlinear benchmark.

\begin{table}[t]
\renewcommand{\arraystretch}{1.12}
\caption{Controlled-intervention design and evidence roles.}
\label{tab:protocol}
\centering
\begin{tabular}{|p{0.8cm}|p{2.2cm}|p{1.9cm}|p{2.1cm}|}
\hline
\textbf{Block} & \textbf{Held Constant} & \textbf{Manipulated} & \textbf{Primary Outputs} \\
\hline
CI-1 & $\Lambda$, $\rho(\Acl)$, $G$, $W$ & Eigenvector geometry / $\kappaV$ & $G_{\mathrm{peak}}$, $\tr(\Sigma_x)$, empirical covariance \\
\hline
CI-2 & $\Acl$, $\rho(\Acl)$, $\kappaV$, $G_{\mathrm{peak}}$ & White vs.\ filtered input & Applied-input variance, $\tr(\hat{\Sigma}_x)$, $J_{\mathrm{peak}}$ \\
\hline
CI-3 & Raw controller and scenario dynamics & White vs.\ filtered injection & Action-var. reduction, covariance reduction, consistency \\
\hline
\end{tabular}
\end{table}

For CI-1, the acceptance criterion is that covariance and peak gain increase systematically with non-normality while spectral radius remains fixed. For CI-2, the acceptance criterion is that source-side suppression lowers applied-input variance and state covariance without changing structural peak gain. For CI-3, the criterion is consistency rather than strict identification: the same source-side intervention should remain beneficial across several task variants under fixed raw closed-loop dynamics. Unless stated otherwise, the synthetic experiments use horizon $T=80$ and $512$ Monte Carlo rollouts. In CI-1, the similarity-shear parameter is swept over $21$ evenly spaced values in $\alpha \in [0,10]$.

\subsection{CI-1: Amplifier Isolation}

We first isolate the amplifier effect using a synthetic family of discrete-time systems
\begin{equation}
\Acl(\alpha) = S(\alpha)\Lambda S(\alpha)^{-1},
\label{eq:synthetic_family}
\end{equation}
with fixed eigenvalues, fixed spectral radius, fixed disturbance channel, and fixed white disturbance covariance. Only the similarity transform $S(\alpha)$ is varied, thereby changing eigenvector geometry and $\kappaV$ while preserving nominal stability.

\begin{figure}[t]
\centering
\subfloat[Covariance amplification at fixed spectral radius.]{
\includegraphics[width=0.48\linewidth]{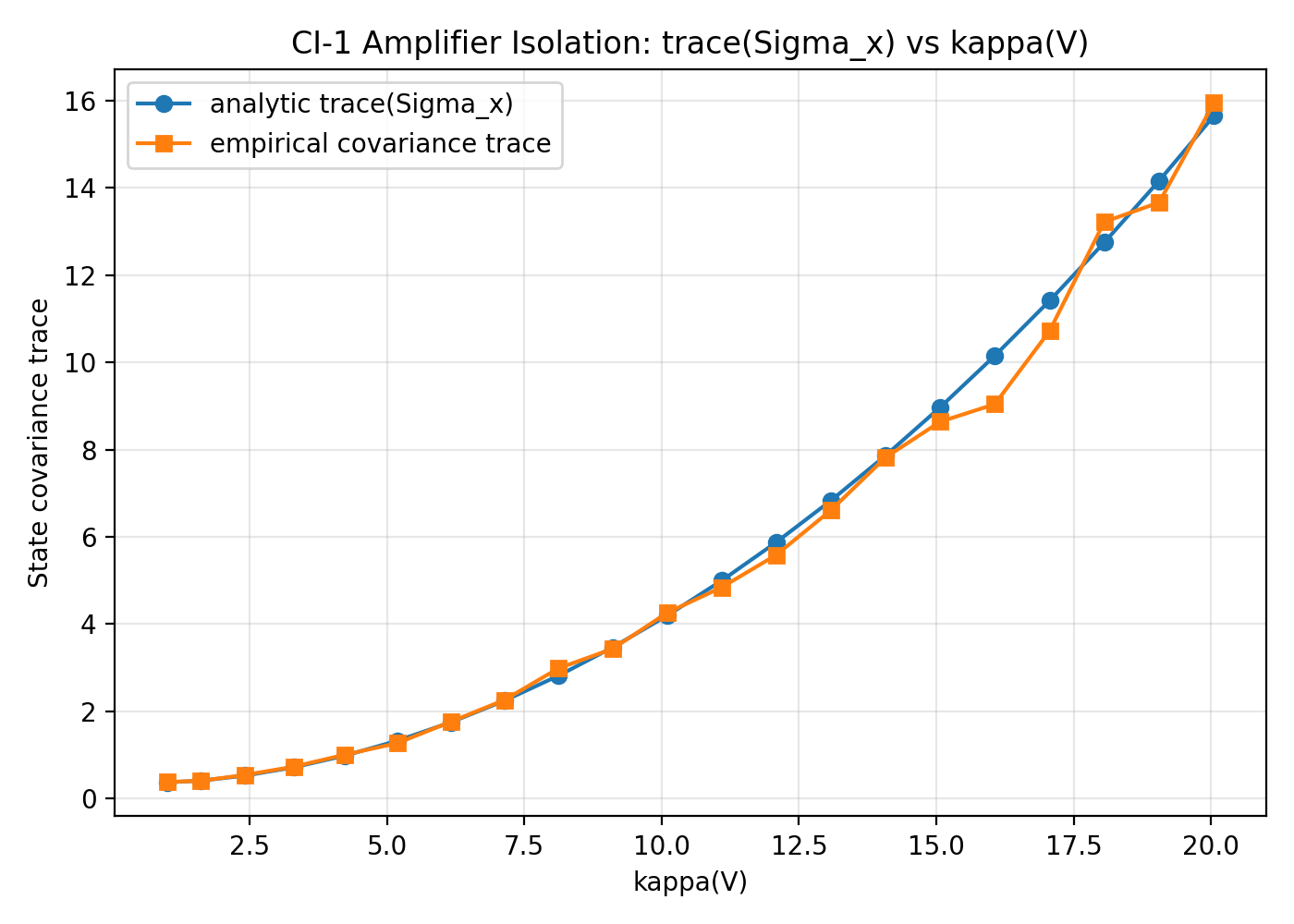}
\label{fig:ci1a}}
\hfil
\subfloat[Peak transient gain at fixed spectral radius.]{
\includegraphics[width=0.48\linewidth]{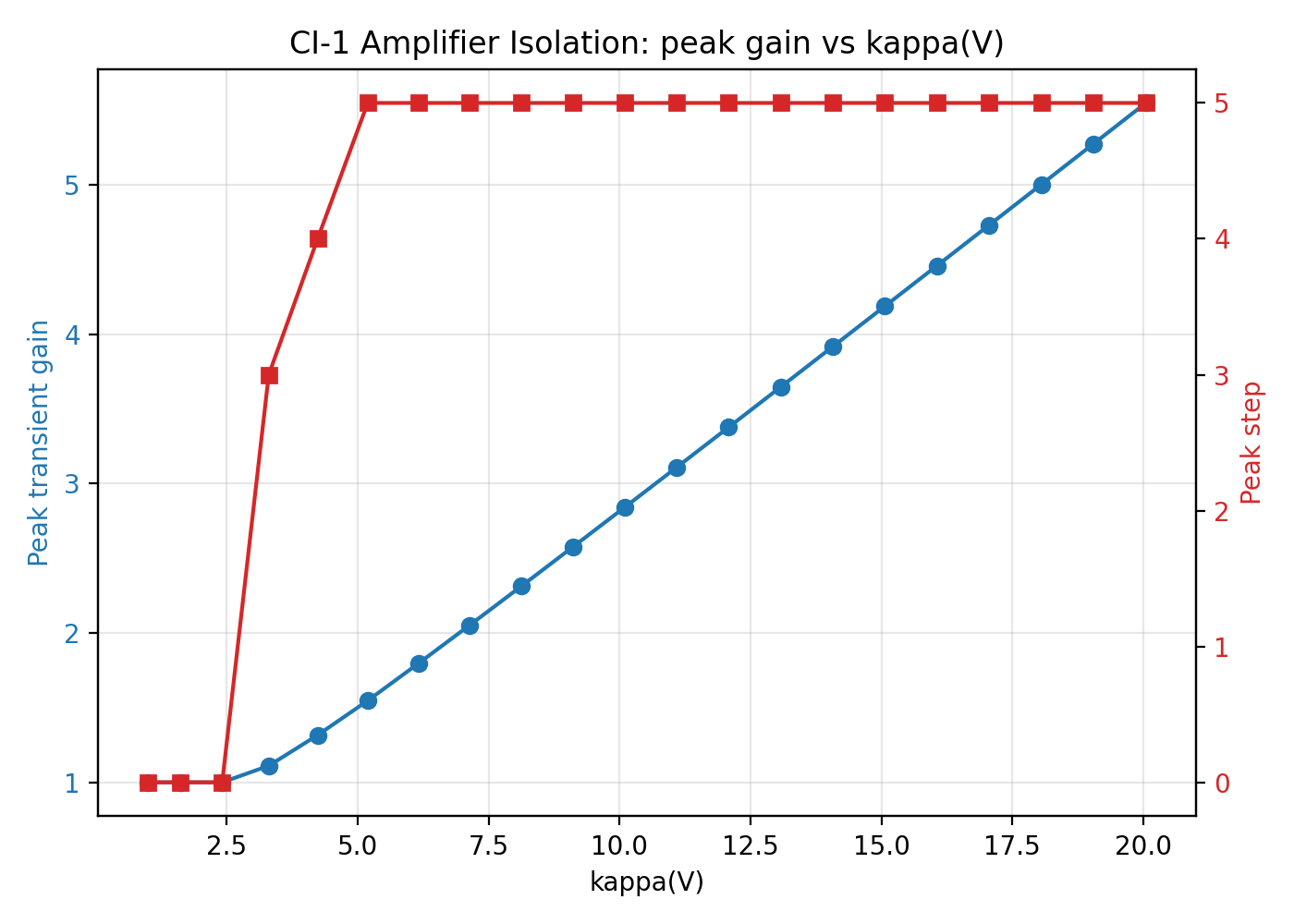}
\label{fig:ci1b}}
\caption{Source--amplifier view of closed-loop variance and amplifier-isolation evidence under fixed eigenvalues, spectral radius, input channel, and disturbance statistics. Increasing non-normality alone raises peak transient gain and state covariance.}
\label{fig:ci1}
\end{figure}

Fig.~\ref{fig:ci1} reports the result. At fixed $\rho(\Acl)=0.93$, increasing $\kappaV$ monotonically raises both $\tr(\Sigma_x)$ and $G_{\mathrm{peak}}$. The observed Pearson correlations are $0.970$ for $\kappaV$ versus covariance trace and $0.996$ for $\kappaV$ versus peak gain. This is the main mechanism-isolation evidence that non-normality can be experimentally isolated as an amplifier.

\begin{table}[t]
\renewcommand{\arraystretch}{1.08}
\caption{Key numerical results for CI-1 and CI-2.}
\label{tab:key_results}
\centering
\footnotesize
\begin{tabular}{|p{2.25cm}|p{1.45cm}|p{2.75cm}|}
\hline
\textbf{Metric} & \textbf{Value} & \textbf{Interpretation} \\
\hline
CI-1: corr.$(\kappaV,\tr(\Sigma_x))$ & $0.970$ & Strong covariance growth under fixed $\rho(\Acl)$ \\
\hline
CI-1: 95\% CI & $[0.958, 0.985]$ & High correlation under bootstrap uncertainty \\
\hline
CI-1: corr.$(\kappaV,G_{\mathrm{peak}})$ & $0.996$ & Peak gain is almost entirely explained by geometry \\
\hline
CI-1: 95\% CI & $[0.993, 1.000)$ & Near-monotone gain growth across the sweep \\
\hline
CI-2: action variance & $0.0399 \rightarrow 0.00317$ & $92.0\%$ reduction under fixed amplifier \\
\hline
CI-2: state covariance & $15.85 \rightarrow 12.39$ & $21.8\%$ reduction with unchanged structural gain \\
\hline
CI-2: expected peak norm & $7.42 \rightarrow 5.56$ & Lower rollout-level state excursions \\
\hline
CI-2: structural peak gain & $5.55 \rightarrow 5.55$ & No change in the structural amplifier \\
\hline
\end{tabular}
\end{table}

\subsection{CI-2: Source-Only Intervention}

We next fix a strongly non-normal closed-loop geometry and compare two source-side conditions: white applied input and filtered applied input. The structural quantities $\rho(\Acl)$, $\kappaV$, and $G_{\mathrm{peak}}$ are held fixed. Only applied-input statistics are changed.

\begin{figure}[t]
\centering
\includegraphics[width=\linewidth]{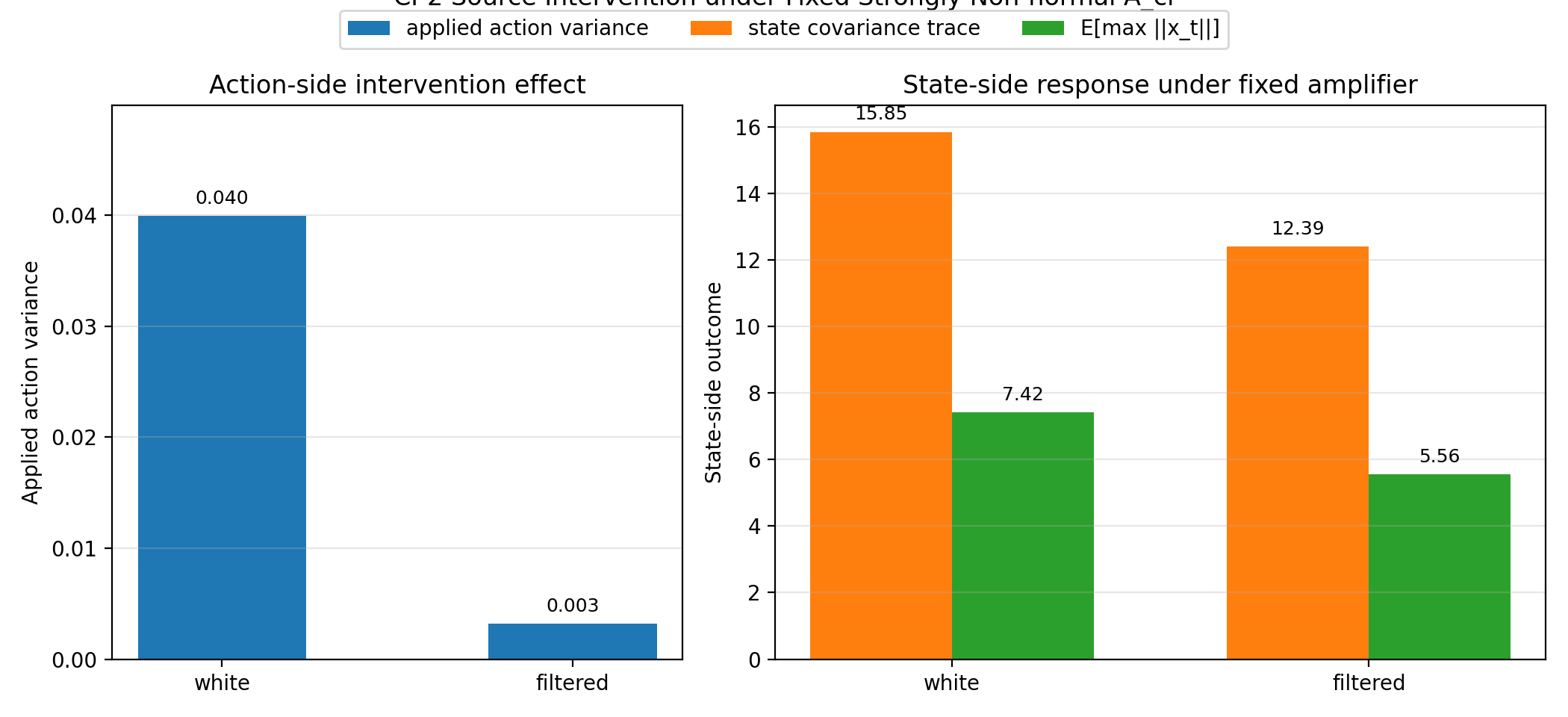}
\caption{Source-only intervention under fixed strongly non-normal dynamics. Input-side suppression reduces applied-input variance and downstream state covariance while leaving structural peak gain unchanged.}
\label{fig:ci2}
\end{figure}

The result is shown in Fig.~\ref{fig:ci2} and Table~\ref{tab:key_results}. Applied-input variance drops from $0.0399$ to $0.00317$, corresponding to a $92.0\%$ reduction. At the same time, state covariance trace drops from $15.85$ to $12.39$, a $21.8\%$ reduction, and the expected peak state norm drops from $7.42$ to $5.56$. Crucially, the structural peak gain remains unchanged by construction. This isolates source-side suppression as a mechanism that reduces state covariance without modifying the amplifier.

\subsection{CI-3: Mechanism-Consistent External Validation on a Planar Quadrotor}

Finally, we bridge the mechanism to a planar quadrotor suite with nominal hover, heavy payload, agile pitch, and payload mismatch scenarios. Here the raw closed loop is kept as the physical branch, while the projected branch remains a surrogate-only analysis/certification branch.

The bridge protocol mirrors the source-side intervention logic of CI-2. For each scenario, we evaluate three random seeds across four disturbance levels $\{0, 0.05, 0.10, 0.20\}$ and compare white versus temporally filtered input perturbations under the same raw controller. Each seed uses $48$ rollouts with horizon $80$. The reported quantities include selected state RMS metrics, full-state RMS, control RMS, empirical covariance trace, and applied-action variance. This experiment is therefore not another causal-isolation stage; instead, it asks whether the mechanism identified in CI-1 and CI-2 continues to organize the observed behavior in a realistic nonlinear control benchmark.

\begin{table}[t]
\renewcommand{\arraystretch}{1.1}
\caption{Planar-quadrotor bridge scenarios and evaluation protocol.}
\label{tab:uav_scenarios}
\centering
\begin{tabular}{|p{2.1cm}|p{4.6cm}|}
\hline
\textbf{Scenario} & \textbf{Role in bridge evaluation} \\
\hline
Nominal hover & Baseline operating point with matched dynamics \\
\hline
Heavy payload & Mass/inertia shift with slower translational dynamics \\
\hline
Agile pitch & More aggressive attitude response with stronger transient coupling \\
\hline
Payload mismatch & Model mismatch setting used to test mechanism consistency under parameter error \\
\hline
\end{tabular}
\end{table}

\begin{figure}[t]
\centering
\includegraphics[width=\linewidth]{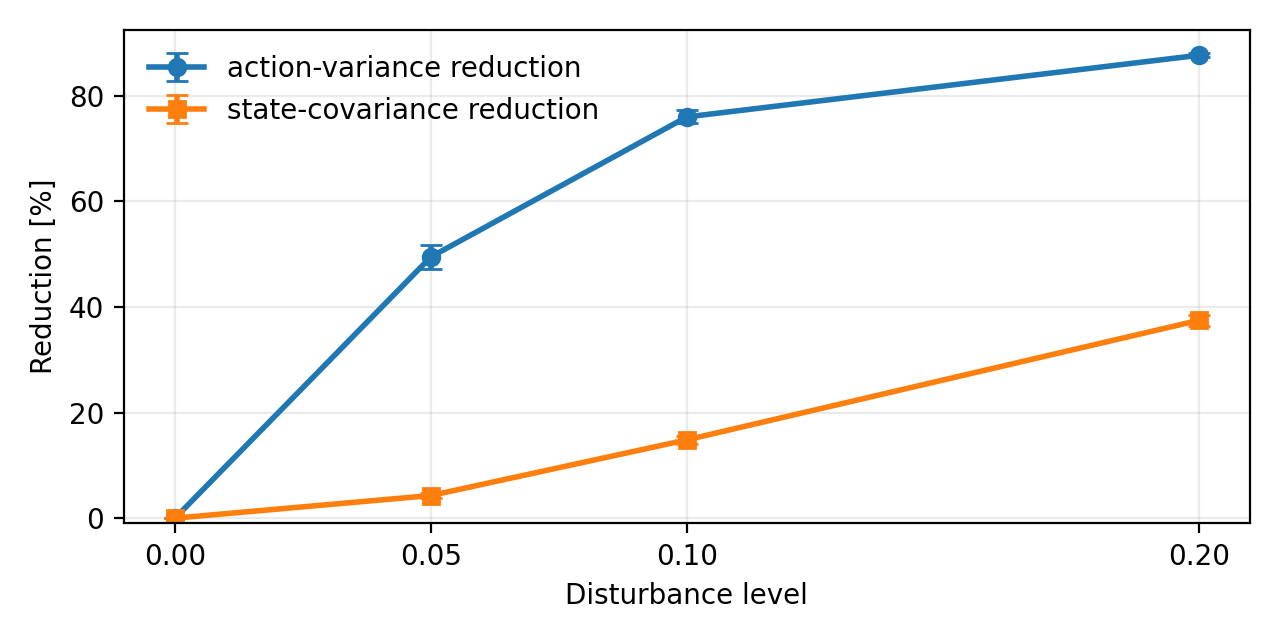}
\caption{Mechanism-consistent external validation on planar quadrotor tasks. Error bars denote standard deviation across three seeds after aggregating reductions over the four scenarios. The UAV bridge evaluates disturbance levels $\{0.00,0.05,0.10,0.20\}$, while the $0.20$ table is retained separately as a compact stress-case snapshot rather than because lower-noise runs were omitted.}
\label{fig:ci3}
\end{figure}

Fig.~\ref{fig:ci3} reports the bridge summary. Aggregating across scenarios and then across seeds, mean action-variance reductions are $49.49 \pm 2.28\%$, $76.00 \pm 1.25\%$, and $87.67 \pm 0.38\%$ at disturbance levels $0.05$, $0.10$, and $0.20$, respectively; the corresponding state-covariance reductions are $4.29 \pm 0.48\%$, $14.84 \pm 0.78\%$, and $37.47 \pm 0.98\%$ (the $0.00$ case is $0$ by construction). At the stress case $0.20$, the raw operating conditions exhibit elevated peak transient responses with mean peak gain $9.28$. This gain summary is obtained from the raw-branch local analysis operator (\texttt{transient\_raw}) associated with each UAV scenario and is reported only as an amplification diagnostic; the covariance and action-variance reductions themselves are measured on the raw nonlinear rollout branch. We interpret these results only as mechanism-consistent external validation: they show that the source-side intervention remains informative in the nonlinear benchmark, not that amplifier isolation has been re-established in the UAV setting.

\begin{table}[t]
\renewcommand{\arraystretch}{1.08}
\caption{High-noise ($0.20$) scenario-level bridge results, retained as a compact stress-case snapshot within the broader four-noise CI-3 evaluation. Reductions are reported as mean $\pm$ standard deviation across seeds.}
\label{tab:uav_numeric}
\centering
\footnotesize
\begin{tabular}{|l|c|c|c|}
\hline
\textbf{Scen.} & $G_{\mathrm{peak}}$ & $\Delta$ act. [\%] & $\Delta$ cov. [\%] \\
\hline
NH & 9.84 & $89.77 \pm 0.34$ & $52.55 \pm 0.64$ \\
\hline
HP & 9.65 & $86.51 \pm 0.92$ & $25.84 \pm 0.75$ \\
\hline
AP & 10.13 & $89.23 \pm 0.55$ & $48.18 \pm 2.69$ \\
\hline
PM & 7.51 & $85.16 \pm 0.27$ & $23.30 \pm 1.55$ \\
\hline
Agg. & 9.28 & $87.67 \pm 0.38$ & $37.47 \pm 0.98$ \\
\hline
\end{tabular}
\end{table}

\subsection{Discussion}

Two questions are worth addressing directly. First, this is not merely an action-smoothing result. CI-2 holds the amplifier fixed and still produces lower downstream covariance, showing that the source-side intervention has a distinct dynamical effect. Second, non-normality is not used merely as a post hoc explanation. CI-1 manipulates amplifier geometry alone and systematically raises covariance and peak gain while nominal stability remains fixed.

For the UAV bridge, all four disturbance levels, $\{0.00,0.05,0.10,0.20\}$, are evaluated over three seeds with $48$ rollouts per seed and horizon $80$. The cross-noise view establishes trend consistency, whereas the $0.20$ case is retained as a compact stress-case snapshot. Unless otherwise noted, CI-3 statistics are reported as mean $\pm$ standard deviation across seeds. This bridge is intended as mechanism-consistent external validation rather than strict causal identification, so it should be read as consistency evidence for the source-side intervention rather than as amplifier isolation.

The intended scope is correspondingly narrow. This letter is complementary to smooth-control RL baselines such as action regularization, smooth exploration, and autoregressive policies rather than a head-to-head benchmark against them. The present results isolate a source--amplifier mechanism that such methods may also implicitly act upon. Likewise, no $\beta$ sweep or multi-task/multi-algorithm benchmark is included in this version. Those broader trade-off and benchmarking questions are left to future work so that the manuscript can remain focused on mechanism isolation and controlled intervention evidence.

\section{Conclusion}

This letter supports a mechanism-focused claim: in the studied continuous-control RL settings, execution-time closed-loop variance is not explained solely by disturbance sources, but can also be materially shaped by non-normal transient amplification. The controlled interventions isolate amplifier-side and source-side effects separately, while the planar quadrotor study provides mechanism-consistent external validation rather than strict causal identification. Throughout, Koopman/ALE surrogates are limited to analysis and certification roles. We therefore position the contribution as an under-emphasized control-theoretic perspective with direct isolation evidence, rather than as a broadly established general account of RL variance behavior.

\section*{Acknowledgment}

The author acknowledges the use of internal simulation and analysis tooling developed for this study.

\bibliographystyle{IEEEtran}
\bibliography{IEEEabrv,paper_lcss_nonnormal_variance}

\end{document}